\documentclass[sigconf]{acmart}

\usepackage{color}
\usepackage{enumitem}
\usepackage{CJKutf8}
\usepackage{ifthen}
\usepackage[normalem]{ulem}
\usepackage{soul}
\usepackage{booktabs}
\usepackage{multirow}
\usepackage{tabularx}
\usepackage{dirtytalk}
\usepackage{url}
\usepackage{breakcites} 
\usepackage{graphicx}
\usepackage{caption}
\usepackage{subcaption}
\usepackage{makecell}
\usepackage{booktabs}
\usepackage{multirow}
\usepackage{amsmath}

\definecolor{gray}{rgb}{0.8,0.8,0.8} 
\definecolor{green}{rgb}{0, 0.6, 0} 
\definecolor{orange}{rgb}{1, 0.5, 0} 	
\definecolor{mahogany}{rgb}{0.75, 0.25, 0.0}
\definecolor{purple}{rgb}{0.6, 0, 0.6}
\definecolor{darkgreen}{rgb}{0, 0.4, 0}
\definecolor{lightblue}{rgb}{0, 0.8, 1} 
\definecolor{lightgreen}{rgb}{0.9, 1, 0.9} 
\definecolor{lightred}{rgb}{1, 0.3, 0.3} 
\definecolor{navy}{rgb}{0, 0, 0.4}
\definecolor{highlight}{rgb}{1.0, 0.13, 0.32}
\definecolor{black}{rgb}{0.0, 0.0, 0.0}

\provideboolean{revising}
\setboolean{revising}{false}

\ifthenelse{\boolean{revising}}
{
    \newcommand{\revision}[1]{\textcolor{blue}{#1}}
    \newcommand{\delete}[1]{\textcolor{gray}{\sout{#1}}} 
    \newcommand{\deletesection}[1]{\textcolor{gray}{{#1}}}

}{
	\newcommand{\revision}[1]{#1}
	\newcommand{\delete}[1]{}
        \newcommand{\deletesection}[1]{}

    \newcommand{\tc}[1]{}
        
}

\AtBeginDocument{%
  }

\copyrightyear{2026}
\acmYear{2026}
\setcopyright{cc}
\setcctype{by}
\acmConference[IUI '26]{31st International Conference on Intelligent User Interfaces}{March 23--26, 2026}{Paphos, Cyprus}
\acmBooktitle{31st International Conference on Intelligent User Interfaces (IUI '26), March 23--26, 2026, Paphos, Cyprus}
\acmPrice{}
\acmDOI{10.1145/3742413.3789105}
\acmISBN{979-8-4007-1984-4/2026/03}

\begin{document}

\title{Key Considerations for Domain Expert Involvement in LLM Design and Evaluation: An Ethnographic Study}

%
\author{Annalisa Szymanski}
\affiliation{
    \department{Computer Science and Engineering}
  \institution{University of Notre Dame}
  \city{Notre Dame}
  \state{IN}
  \country{USA}
}
\email{aszyman2@nd.edu}

\author{Oghenemaro Anuyah}
\affiliation{%
 \institution{Microsoft Corporation}
 \city{Redmond}
 \state{WA}
 \country{USA}
 }
\email{maroanuyah@microsoft.com}

\author{Toby Jia-Jun Li}
\email{toby.j.li@nd.edu}
\affiliation{
\department{Computer Science and Engineering}
\institution{University of Notre Dame}
\city{Notre Dame}
\state{IN}
\country{USA}}

\author{Ronald A. Metoyer}
\affiliation{
\department{Computer Science and Engineering}
\institution{University of Notre Dame}
\city{Notre Dame}
\state{IN}
\country{USA}}
\email{rmetoyer@nd.edu}

%
\renewcommand{\shortauthors}{Szymanski et al.}

\begin{abstract}
Large Language Models (LLMs) are increasingly developed for use in complex professional domains, yet little is known about how teams design and evaluate these systems in practice. This paper examines the challenges and trade-offs in LLM development through a 12-week ethnographic study of a team building a pedagogical chatbot. The researcher observed design and evaluation activities and conducted interviews with both developers and domain experts. Analysis revealed four key practices: creating workarounds for data collection, turning to augmentation when expert input was limited, co-developing evaluation criteria with experts, and adopting hybrid expert–developer–LLM evaluation strategies. These practices show how teams made strategic decisions under constraints and demonstrate the central role of domain expertise in shaping the system. Challenges included expert motivation and trust, difficulties structuring participatory design, and questions around ownership and integration of expert knowledge. We propose design opportunities for future LLM development workflows that emphasize AI literacy, transparent consent, and frameworks recognizing evolving expert roles.
\end{abstract}




\maketitle

\section{Introduction}

Large Language Models (LLMs) are increasingly used in high-stakes domains such as education, healthcare, and law, where the accuracy of outputs carries important ethical and practical implications~\cite{sarkar2024llms, sallam2023chatgpt, ahmad2023creating, chatelan2023chatgpt, ayers2023comparing}. These models can produce fluent, contextually relevant responses that expand access to expertise and assist in complex decision-making~\cite{tian2023opportunities, benary2023leveraging}. However, these same models may also reproduce bias, generate misleading outputs, or hallucinate information~\cite{ji2023survey, huang2023survey}. As LLMs become more accessible through open-source and commercial platforms, questions of validation, reliability, and appropriate deployment have become increasingly urgent~\cite{manchanda2024open, hou2024large, zhao2023survey}. 
\revision{The reliability of LLM-based systems depends not only on technical performance but also on how development teams curate data, define fine-tuning objectives, and establish evaluation criteria~\cite{qi2023fine, barnett2024fine}, requiring teams to balance scalability, rigor, and contextual relevance across iterative development cycles~\cite{liusie2024efficient}.} 
\delete{The reliability of LLM-based systems ultimately depends not only on technical performance but also on how development teams design and evaluate their systems. Design choices made during model development, such as how training data are curated, fine-tuning objectives are defined, and evaluation criteria are established, affect both model behavior and the standards by which they are evaluated~\cite{qi2023fine, barnett2024fine}. LLM development teams must make a series of decisions about how to balance scalability, rigor, and contextual relevance across iterative cycles of model refinement and evaluation~\cite{liusie2024efficient}.} 

Several strategies for evaluating LLMs have emerged, including performance benchmarks~\cite{papineni2002bleu, lin2004rouge}, LLM-as-a-Judge methods~\cite{zheng2024judging, li2023alpacaeval, dubois2024alpacafarm}, and direct review by domain experts~\cite{sun2024lalaeval, pan2024human}. Each approach offers trade-offs: benchmarks and automated metrics may fail to capture domain-specific requirements; LLM-as-a-Judge methods, though scalable, can introduce bias or overlook subtle errors~\cite{szymanski2025limitations, chen2024humans}; and domain expert evaluation can provide rich, contextualized feedback~\cite{vo2021role, michael2019clinical, martin1999influences}, but their involvement is resource-intensive, time-consuming, and often limited~\cite{gehrmann2023repairing}. While these techniques are often studied in isolation, little is known about how they are combined in real-world development workflows, how teams involve experts, or how organizational factors influence these decisions.

Unlike traditional software, LLM-based systems are non-deterministic and require human judgment and iterative refinement~\cite{brown2020language, song2024good, barr2014oracle}. Development involves coordination across technical and domain expertise, often under constraints of time, interpretability, and institutional expectations. Yet, current research offers limited insight into how these trade-offs are negotiated \textit{in situ} and how reliability is prioritized in development work. To address this gap, we argue for empirical approaches that examine how evaluation and design decisions are made in practice. Ethnographic methods allow researchers to observe not only formal workflows but also informal practices, negotiations, and constraints that shape design and evaluation~\cite{geertz2017interpretation, saxena2021framework}. By embedding in the team’s day-to-day work, we aim to understand how decisions about expert involvement and evaluation are made, and what trade-offs emerge. This work addresses the following research questions:



\begin{itemize}
    \item \textbf{RQ1:} How do members of an LLM development team carry out design and evaluation activities over the course of a development cycle?
    \item \textbf{RQ2:} What challenges and trade-offs emerge in decision-making about evaluation and expert involvement?
    \item \textbf{RQ3:} What lessons can inform better workflows, tools, and practices for supporting domain experts in LLM development?
\end{itemize}



\revision{We conducted a 12-week participatory ethnographic study of a university-based team developing an LLM-powered pedagogical chatbot.} 
Through \revision{a thematic analysis of} observations and interviews with both developers and pedagogy experts, we found that the team approached design and evaluation through four interrelated practices: (1) creating workarounds to collect gold-standard conversations from experts, (2) turning to data augmentation when expert data was limited, (3) utilizing collaboration with experts to co-develop evaluation criteria, and (4) adopting hybrid evaluation strategies that combined expert input, developer judgment, and LLM-as-a-Judge scoring. These practices illustrate how the team balanced rigor with time and resource constraints. However, expert involvement also introduced challenges. Experts reported fatigue with repetitive tasks, struggled with limited AI literacy when asked to design evaluation criteria, and expressed concerns about embedding their knowledge in training data. Despite these challenges, expert involvement was motivated by institutional trust, alignment with the project’s pedagogical mission, and the opportunity to shape a tool for their community.

We contribute insights into the design and evaluation practices that evolve during LLM development, the constraints that shape expert involvement, and the requirements to sustain meaningful expert engagement over time. We summarize the following research contributions:

\begin{itemize}
    \item An in-depth understanding of the design and evaluation decisions made by an LLM development team during the creation of a domain-specific chatbot.
    \item An analysis of the challenges and trade-offs involved in integrating domain expertise, drawn from both team observations and interviews.
    \item Key considerations for promoting expert–developer collaboration, including ethical implications and best practices for involving domain experts in LLM development.
\end{itemize}

\section{Background and Related Work}

\subsection{Existing Approaches to Studying LLM Development}

To gain insight into LLM development, previous literature has used surveys, interviews, and large-scale data analyses. Studies of public developer communities have characterized challenges in prompt design, API use, and plugin integration~\cite{chen2025empirical}, as well as fine-tuning, dataset management, and deployment~\cite{alam2024developer}. Analyses of open-source software have identified model, parameter, and integration challenges in generative AI applications~\cite{cai2025demystifying, shao2024llms}. Survey-based studies have mapped methods, metrics, and benchmarks for LLM tasks such as model compression~\cite{zhu2024challenges} and evaluation~\cite{peng2024survey, chang2024survey, gu2024survey, cao2025toward}, while qualitative work has examined issues of bias, privacy, and usability~\cite{jalil2025transformative, ruan2025qualitative, pasaribu2024development}. Although these studies provide valuable overviews, most rely on retrospective or aggregated data and give limited visibility into how development decisions are made in context~\cite{park2024understanding, nahar2025beyond}. 

As a result, we know little about the day-to-day work through which teams interpret model behavior, negotiate evaluation strategies, and adapt to constraints. Ethnographic methods, by contrast, capture these practices as they unfold to reveal what practitioners do and why they make particular decisions~\cite{sharp2016role, passos2012challenges}. This approach provides a critical complement to existing survey- and interview-based studies by grounding observations of development work.

\subsection{Integrating Domain Experts in LLM Design}
Prior research in HCI has emphasized the value of incorporating multiple stakeholders in AI development, yet less attention has been given to how development teams decide when and how to involve them within the constraints of organizational context. In applied machine learning (ML), domain expertise often enters the pipeline through the creation of gold-standard datasets, where experts label or provide examples that serve as benchmarks for training and evaluation~\cite{gururangan2020don}. This approach is common in fields such as healthcare and law, where expert judgment defines correctness~\cite{vo2021role}. However, relying on static datasets limits adaptability when models are deployed in dynamic professional settings and requires significant time and resources from experts~\cite{borger2023artificial, gehrmann2023repairing}.

Beyond dataset construction, participatory and co-design methods have emerged to embed expertise more directly into the design process. Human-in-the-loop approaches enable experts to guide requirement gathering, prompt design, and model evaluation~\cite{reza2025prompthive, szymanski2024integrating, shah2025prompt}. Participatory workshops have also been used to co-create evaluation rubrics, elicit tacit knowledge, and ground model outputs in professional standards~\cite{pan2024human, sun2024lalaeval}.  However, while participatory methods capture context-specific expertise that benchmarks often miss, they are difficult to scale and require substantial coordination and time commitment from participants~\cite{gehrmann2023repairing}. 

Despite these advances, little is known about how expertise is integrated into day-to-day development. Most studies focus on discrete activities such as annotation or evaluation, without examining how teams balance expert input, automated methods, and shifting project goals.


\subsection{Strategies for Evaluating LLMs}

Evaluating LLMs remains an active area of research, especially for complex tasks where accuracy and contextual understanding are difficult to quantify~\cite{chang2024survey}. Traditional automated metrics such as BLEU and ROUGE~\cite{papineni2002bleu, lin2004rouge} offer scalability but often miss the domain-specific qualities that matter most in applied settings. As a result, human evaluation has become the preferred standard~\cite{sun2024lalaeval, pan2024human}.

Human evaluators provide grounded judgments that capture reasoning, context, and appropriateness~\cite{wu2022survey}, but their involvement can be costly and time-consuming~\cite{gehrmann2023repairing}. Expert input is especially valuable in specialized domains to align model outputs with professional standards~\cite{cheng2023now, liu2023chatcounselor}. However, these evaluations are difficult to scale and sustain and motivates interest in automated alternatives. To address these limitations, researchers have explored the LLM-as-a-Judge approach, where a separate model acts as an evaluator to approximate human judgment~\cite{zheng2024judging, li2023alpacaeval, dubois2024alpacafarm}. Under certain conditions, these automated judges show promising correlation with human preferences~\cite{dubois2024length} and offer a scalable complement to human review. Typically, LLM evaluators apply pre-defined natural-language evaluation criteria, or statements reflecting correctness, completeness, clarity, or tone, to rank model responses~\cite{shankar2024validates, zhang2024llmeval}. 

In real-world contexts, these LLM-mediated evaluations offer several practical benefits. They help developers identify strengths and weaknesses in model performance, detect emerging errors, and decide when to refine prompts or retrain models~\cite{arawjo2023chainforge, shankar2024validates}. They can also serve as early indicators of when human or expert input may be necessary to support more efficient allocation of limited human review. Continuous, criterion-based evaluation~\revision{\cite{shankar2024validates, desmond2024evalullm}} may help developers and end-users build confidence in LLM outputs to offer guidance on when systems can be trusted or require further oversight~\cite{liu2023trustworthy}. However, LLM evaluators inherit the same biases and limitations as the models they assess\revision{~\cite{zheng2024judging, shi2024judging, chen2024humans}}, which further complicates decisions about when their judgments can be trusted and when expert oversight is needed. 

Understanding how development teams navigate these questions, how they decide when to rely on automated evaluation, when to seek expert input, and how to reconcile the two, is critical for building responsible, reliable LLM systems. Through an ethnographic lens, we trace how these methods were integrated, adapted, and negotiated throughout the creation of a pedagogical chatbot to reveal how evaluation workflows evolve under real-world constraints.
\section{Method}
\delete{The goal of this study is to investigate how members of an LLM development team design and evaluate their system, what challenges and trade-offs they face, and how their practices might inform future workflows that support domain experts in LLM design. To address our research questions (RQ1-RQ3), we conducted a 12-week ethnographic study. This section outlines our ethnographic rationale, study context, participants, data collection procedures, ethical considerations, and analytic process.}

\subsection{Ethnographic Approach and Rationale}

\revision{To address our research questions (RQ1-RQ3), a 12-week ethnographic study was conducted.} We adopted a participatory ethnographic approach to obtain a situated understanding of day-to-day practices, interactions, and meanings that shaped design and evaluation decisions within the development team. Ethnography enables the production of thick description \cite{geertz2017interpretation, saxena2021framework}, revealing how values, constraints, and professional expertise are negotiated through practice. The first author, referred to as the researcher, served as a member of the development team, engaging in meetings and informal discussions while maintaining a reflective and analytical stance. This approach allowed for direct access to how design and evaluation unfolded in real-time and revealed both explicit decisions and the reasoning behind them.

The ethnographic method was chosen for two main reasons. First, LLM development differs from traditional software engineering in its iterative and interpretive nature: model behavior is unpredictable, evaluation criteria evolve dynamically, and expert input is often negotiated under resource and organizational constraints. Ethnography captures how these indeterminate processes unfold \textit{in situ} and reveals the tacit judgments that shape the technology. Second, by participating as a recognized member of the team while maintaining an analytic stance, the researcher could access both formal meetings and informal exchanges central to sense-making. This positioning allowed the study to trace how developers and pedagogical experts defined ``good'' performance, justified evaluation decisions, and established boundaries between human and LLM responsibility.

\subsection{Study Context and Participants}
This study was conducted between June and August 2025 with a development team at a private university developing an LLM-powered pedagogical chatbot. The project took place within the university’s teaching and learning center, an academic unit providing pedagogical resources, consultations, and programming for faculty and instructors. The center’s mission is to promote effective teaching practices and enhance learning by offering instructors individualized consulting sessions, workshops, and ongoing professional development opportunities. The consultants, or domain experts, are experienced educators who support instructors by providing expertise in pedagogy, instructional design, and faculty development. The chatbot project, referred to as \textit{InstructAI}, originated from a collaboration between the center and a development team within the institution. The goal was to create an AI-powered system that provides real-time, pedagogically grounded teaching suggestions to faculty. Specifically, InstructAI was envisioned as a tool that could respond to instructors’ questions about classroom challenges, teaching strategies, or student engagement by drawing upon a knowledge base of best practices and pedagogical guidance.

\begin{figure*}[!t]
\centering
\includegraphics[width=\textwidth]{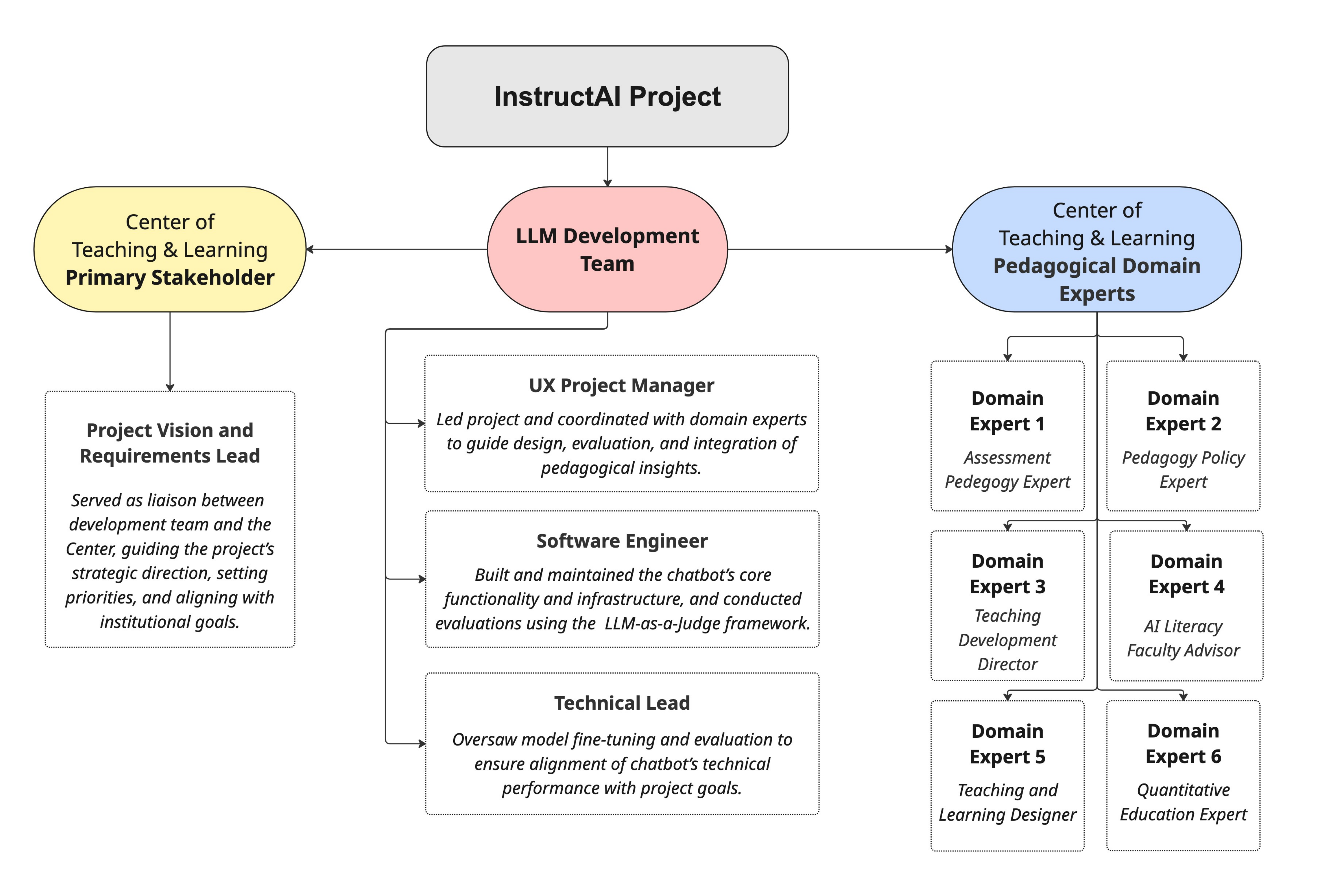}
\caption{\revision{\textbf{Stakeholder structure and roles in the InstructAI project.} The LLM development team coordinated design, implemented the system, and oversaw model development and evaluation. The team worked in collaboration with a primary stakeholder from the center of teaching and learning, who set the project vision and requirements, and with pedagogical domain experts, who contributed expertise during design and evaluation activities.}}
\label{fig:stakeholders}
\end{figure*}

When this study began, the project was entering a critical phase of transitioning from initial requirements gathering to iterative model fine-tuning, evaluation, and system testing. Prior to the researcher joining, the development team had interviewed several pedagogical consultants and facilitated an early design workshop to storyboard potential chatbot features and was ready to integrate domain experts into the evaluation loop and begin systematic testing of the model’s responses. The development team operated semi-independently from the center but maintained close collaboration through recurring joint sessions. \revision{It consisted of three members (UX project manager, technical lead, and software engineer) that operated within a U.S.-based academic institution, all with prior experience in AI and LLM research or development. The UX project manager had prior training in AI in education and human–computer interaction, the technical lead brought prior industry research experience and expertise in developing and optimizing large language models, and the software engineer contributed a background in full-stack web development and machine learning, including experience integrating generative AI components into interactive systems.} 
The development team worked closely with a primary stakeholder, a pedological scholar for the center, responsible for setting the project vision and coordinating requirements. 
The domain experts who were affiliated with the  center, participated in design and evaluation activities at key points, such as co-developing evaluation rubrics, reviewing chatbot outputs, and advising on pedagogical framing. Figure \ref{fig:stakeholders} depicts this multi-stakeholder structure, with both technical contributors and pedagogical specialists.


\subsection{Researcher Role and Participatory Ethnographic Approach}


This study employed a participatory ethnographic approach, positioning the first author as an embedded researcher within the LLM development team. Unlike a detached observer, participatory ethnography involves active engagement in the social setting while maintaining reflexive awareness of the researcher’s dual role as a participant and analyst \cite{madden2022being}. \delete{This approach was particularly well suited for examining how design and evaluation decisions evolved \textit{in situ}, where much of the meaningful negotiation occurred informally, through conversations and everyday workarounds.} As a recognized and trusted member of the team, the researcher participated in daily stand-ups, weekly planning meetings, and evaluation sessions, contributing observations and perspectives when relevant but intentionally refraining from taking ownership of project deliverables or influencing final design decisions. This positioning provided access not only to formal design and evaluation practices but also to the reasoning, trade-offs, and interpersonal dynamics that shaped them.

The researcher had access to all team communications, including Slack channels, shared documents, and email threads, which enabled observation of ongoing interactions where design and evaluation decisions were negotiated. The team’s awareness of the researcher’s dual role encouraged reflective discussions, which allowed observation of collective sensemaking and self-critique. This embedded perspective also captured developer-expert interactions that shaped how expert knowledge was incorporated into LLM evaluation. Observing and occasionally facilitating these engagements revealed how expert contributions were framed, interpreted, and operationalized across iterative development cycles. This participatory lens captured not only what the team did, but how technical and pedagogical members understood, justified, and adapted their practices in response to emerging challenges.

\subsection{Data Collected}


To capture a holistic view of the development and evaluation process, multiple qualitative data sources were drawn upon. These included daily and weekly observations of team practices, interviews with both developers and domain experts, and a collection of artifacts that documented design decisions, communications, and evaluation outcomes.

\subsubsection{Daily Observations}
Over a 12-week period, the researcher conducted 70.8 hours of observations embedded within the team’s day-to-day workflow. The team held daily 30-minute stand-ups (focusing on blockers, evaluation tasks, and model updates) and two recurring weekly meetings (30 minutes and 75 minutes) with the primary stakeholder to align evaluation plans, rubric revisions, and milestone targets. The researcher also observed ad-hoc activities central to the project’s design and evaluation loop: one-on-one strategy check-ins, small-group discussions on emergent issues, planning sessions for upcoming design/evaluation workshops, and both virtual and in-person team checkpoints. Integration into Slack and email channels, allowed observation of informal negotiations where decisions formed, issues emerged, trade-offs were accepted, and priorities shifted with stakeholder input or evaluation results. Observations also extended to sessions with domain experts, including design workshops, rubric co-development, review of AI-generated outputs, and debriefs after both evaluation and data contribution activities (e.g., collaborative sessions to develop evaluation criteria or provide exemplar data for model training). 

The researcher balanced participation with analytic distance by contributing context when invited without owning deliverables or decisions. During meetings, the researcher produced real-time jottings and same-day expanded field notes. Notes were organized under consistent headings: Actors, Activities, Artifacts (documents, prompts, rubrics), Decision Points, Rationales, Tensions/Trade-offs, Follow-ups. Furthermore, the notes were accompanied by weekly analytical memos that captured emerging interpretations. These memos were revisited to check for interpretive drift and to refine questions for subsequent observations or interviews.

Analytically, observation time concentrated on key episodes where design and evaluation intertwined:

\begin{itemize}
    \item Gold-standard data creation (how exemplar responses were elicited from domain experts, prepared for fine-tuning, and positioned as reference material for evaluation).

    \item Evaluation-method design (introducing, revising, or retiring criteria; deciding thresholds; debating inter-rater alignment between developers and pedagogical experts).

   \item Human–automation balance (when to invoke LLM-as-a-Judge vs. expert review; how to resolve disagreements; how automated scores influenced prioritization).

  \item Stakeholder alignment (how pedagogical priorities, timeline constraints, and technical feasibility were reconciled).
\end{itemize}

This structured and flexible observation strategy yielded a thick description of everyday development and evaluation practices to capture both formal processes and the informal workarounds through which the team progressed.

\subsubsection{Interviews}

We combined informal, \textit{in-situ} conversations with semi-structured interviews to investigate situated practices and reflective accounts of the design and evaluation. Informal interviews (contextual inquiries) were conducted continuously after meetings or key episodes (e.g., rubric changes, conflicting scores), while semi-structured sessions were scheduled once the team had completed several evaluation cycles, allowing participants to reflect across iterations. We interviewed all members of the development team (\textit{n} = 3) to capture complementary perspectives across technical leadership, engineering, and research roles. We also interviewed the the primary stakeholder and domain experts from the center who had directly participated in design and/or evaluation activities (e.g., rubric co-development, output review, data collection tasks). Experts were selected via criterion sampling to ensure variation in pedagogical specialization and degree of involvement across the study period.

Each interview session lasted approximately 45 to 60 minutes and was conducted toward the end of the 12-week study, once participants had substantial experience with design and evaluation activities. We used insights from our field notes and early observations to develop two tailored semi-structured interview protocols: one for developers and one for domain experts. The interviews were tailored to deepen understanding of how design and evaluation unfolded in practice and to complement the ongoing participant observation data.

\textbf{Development Team interviews.} Interviews with the three development team members focused on their roles, responsibilities, and reflections on integrating evaluation into the system’s lifecycle. Questions probed how evaluation was built into the development process, whether it was planned from the beginning or added iteratively, and how collaboration with pedagogical experts shaped decision-making. For instance, developers were asked: \textit{``Can you walk me through how evaluation was built into the development process? Were evaluation and feedback considered from the beginning, or integrated later?''} \textit{``In your view, where is human expert input most necessary, and where might an LLM be sufficient?''} and \textit{``What challenges did you encounter when trying to incorporate domain expertise into design or evaluation?''}

These questions were grounded in themes that emerged during daily observations, such as the negotiation between automated evaluation (LLM-as-a-Judge) and human expert review, as well as the trade-offs between model performance goals and pedagogical fidelity. 
Developers were also asked to reflect on tensions between the team’s technical priorities and the center’s educational mission, and to suggest improvements to future workflows or protocols.

\textbf{Domain Expert Interviews.} Interviews with pedagogical consultants (domain experts) examined their experiences participating in the design and evaluation of the chatbot. Experts were asked to describe how they were recruited, what kinds of evaluation tasks they performed, and how clearly those tasks were framed. For example, questions included: \textit{``Can you describe how you were involved in the evaluation or data collection process for InstructAI? What kind of tasks were you asked to complete—reviewing outputs, designing rubrics, or participating in the teacher simulator?''} \textit{``How clear were the instructions or expectations for your role in this process? Did you feel your contributions were taken into account?''} \textit{``Based on your experience, in what areas do you feel your evaluation as an expert added the most value, and where might an LLM do just as well or even better?''} Additional questions invited reflection on comfort and motivation: \textit{``How did you feel about the time and effort the process required?''} and \textit{``What would you want developers of AI systems to understand about the importance of domain expertise?''}

By tailoring questions to participants’ actual engagement in the project, such as their involvement in designing rubrics, reviewing AI outputs, or calibrating evaluation criteria, we encouraged concrete, example-driven accounts rather than abstract opinions. This helped reveal how participants made sense of their contributions, what they found meaningful or challenging, and how they viewed the balance between expert judgment and automated evaluation.

All interviews were audio-recorded and transcribed verbatim, and identifiers were removed to ensure confidentiality. Informal follow-up conversations were also documented in field notes to clarify statements or revisit emerging interpretations. Together, these interviews provided insight into both how design and evaluation practices were enacted and how participants understood and rationalized their roles in shaping the LLM’s development.

\subsubsection{Artifacts}
In addition to observations and interviews, a range of artifacts was collected that documented the team’s iterative design and evaluation process. These included draft evaluation rubrics, participatory design session materials, chatbot outputs, annotated model responses, expert-generated data, and LLM-as-a-Judge results and performance summaries. These materials showed how evaluation criteria evolved, how expert input was incorporated, and how model performance was interpreted over time, which complemented what was observed in meetings and interviews.


\subsection{Ethical Considerations}
This study was approved by the university’s Institutional Review Board (IRB), and all participants provided informed consent. The researcher’s role as an embedded team member was disclosed to all collaborators, and care was taken to maintain voluntary participation and confidentiality. Identifiers were removed during transcription, and pseudonyms are used throughout. Because the researcher was involved in team activities, reflective memos and regular debriefs were used to mitigate bias and ensure analytic transparency. No student or instructional data were used, and all materials were stored on secure, access-restricted servers in accordance with institutional guidelines.

\subsection{Data Analysis}
We conducted a thematic analysis~\cite{braun2006using} across observation field notes and interview transcripts, complemented by an analytic review of collected artifacts. The first author led the analysis and interpretation of the data. 

\revision{For RQ1, analysis focused on longitudinal observation field notes, interviews with developers, and artifacts documenting day-to-day development work. Through repeated readings, the researcher embedded initial codes capturing observable practices (e.g., refining evaluation criteria, coordinating with experts). These codes were iteratively grouped into subthemes (e.g., ethical barriers to real data, expert input improves benchmarks) and then synthesized into higher-level themes describing interrelated design and evaluation practices (e.g., designing workarounds to collect gold-standard conversations, turning to augmentation strategies due to limited data).}
\delete{The process began with iterative readings of field notes and transcripts to identify recurring activities, decisions, and points of negotiation.} Artifacts such as evaluation criteria rubrics, annotated model outputs, and expert evaluation documents were examined to trace the evolution of criteria and to support insights emerging from observations and interviews.

\revision{For RQ2, analysis drew primarily on interviews with developers and pedagogical experts. Interview transcripts were coded for expressed concerns and constraints (e.g., confusion about task focus, loss of ownership), which were iteratively grouped into subthemes (e.g. difficulty understanding evaluation abstractions, concerns about giving away expertise) and higher-level themes (e.g. challenges in structuring participatory design processes, concerns about how expert feedback would be used) capturing key challenges and trade-offs in expert involvement.}

\revision{Across both analyses, themes were refined through repeated engagement with the data, reflexive memo writing, and regular discussions with co-authors. Although formal inter-coder reliability was not employed, these discussions supported critical reflection on analytic decisions, challenged interpretations, and guided the iterative refinement of emerging themes.}


\delete{Initial open codes captured both observable practices (e.g., refining evaluation criteria, coordinating with experts) and underlying rationales (e.g., efficiency, trust, or pedagogical alignment). Codes were iteratively refined into subthemes and higher-level themes that described patterns across the design and evaluation process. The additional authors provided ongoing feedback, helping refine boundaries and ensure analytical coherence. The first author also maintained a reflexive memo journal to document interpretive decisions and positionality, thereby supporting consistency across data sources.}

\subsection{\revision{Positionality Statement}}


\revision{All authors are computer scientists with strong backgrounds in human–computer interaction and human-centered computing. The first author brought prior experience in human–AI interaction research, including extensive work on LLM evaluation and methods for incorporating domain experts into LLM evaluation processes. This background supported informed interpretation of evaluation and design practices during analysis and guided clarifying questions in discussions with developers and domain experts when needed. 
The remaining authors contributed complementary expertise in community-centered technology research, qualitative and ethnographic methods, and technical development and evaluation of LLM systems. Collectively, these perspectives shaped the study’s attention to situated practices, trade-offs, and organizational constraints influencing LLM evaluation. Analytic rigor was supported through reflective memo writing, iterative review of data, and regular collaborative discussions among the research team.}


\section{Findings}

\subsection{Development Team Design and Evaluation Practices \revision{(RQ1)}}

Analysis revealed that the development team carried out design and evaluation activities through four interrelated practices: (1) creating workarounds to collect gold standard conversations from pedagogical experts, (2) turning to augmentation strategies due to limited data collected, (3) utilizing collaboration with pedagogical experts to co-develop evaluation criteria, and (4) adopting hybrid evaluation approaches to address time constraints and inconsistencies. These practices illustrate the adaptive strategies the team employed in the development cycle. 

\subsubsection{Developers Designed Workarounds to Collect Gold Standard Conversations from Pedagogical Experts}
The development team experimented with multiple strategies to collect what they referred to as ``gold standard'' conversational data from pedagogical experts. Their goal was to obtain authentic examples of pedagogical consultation that would serve two purposes: (1) as training data for fine-tuning the model, and (2) as reference data for evaluating model outputs. 
The dataset was intended to capture authentic conversations between instructors and pedagogical consultants that the chatbot was designed to model and to illustrate how a pedagogical expert would respond.

The team’s initial strategy was to secure recordings of real consultations between pedagogical experts and faculty clients that occurred at the university’s teaching and learning center, where consultations were frequent and took place daily. However, a challenge arose regarding the ethical concerns of recording these consultations as part of the data collection and design process. Developers learned that the process of recording could alter the nature of the conversations, discourage faculty participation, and compromise the trust central to these consultations. One developer reflected on the difficulty of securing access:

\begin{quote}
    \textit{``Recording would significantly change the [center's] current services...and probably wouldn’t work...and I wasn’t going to push...they weren't seeing the direct benefit [of recording]. It would require extra work like getting consent...and it might even negatively impact [the consultation] service.''} (UX Project Manager)  
\end{quote}

As a workaround for collecting consultation data, the team recorded a small set of staged conversations in which members of the development team, together with participants who had classroom teaching experience, met with a pedagogical expert for consultation. Developers were limited to recruiting four participants due to time and scheduling constraints. Although these recordings generated some useful material, the developers quickly realized that the dataset was too small and too labor-intensive to clean, since the transcripts contained disfluencies, short filler responses, and language patterns that required extensive editing. They also noted that the conversational style differed from how a user may interact with an AI chatbot, making it less suitable for training or evaluation purposes. 
\delete{Although developers were initially uncertain about what types of data would be most useful, experimentation with recording consultations helped them refine their approach.}

To take a different approach, the team made a pivotal decision to move away from relying on recordings or small staged conversations and instead collected data in text form. For this purpose, developers created a simulation that generated consultation prompts and captured the responses of pedagogical experts in a chatbot-style format. They quickly prototyped the system, training it to adopt an instructor persona, pose an initial question about instruction, and provide follow-up responses. The development team then recruited 18 domain experts and invited them to interact with the chatbot, contributing asynchronously at their own pace and with hourly compensation. Through this strategy, the team collected over 130 simulated dialogues to create a more scalable form of ``gold standard'' data. 

In comparing the two approaches, one developer highlighted why the simulator data felt more practical than transcripts from real conversations:

\begin{quote}
\textit{``[Real conversations] need extra effort to clean by removing those short answers or other fillers that aren’t useful. But chatting [through text] is another good way to collect data, because people may [be inclined to] type more when chatting with [the simulator].''} (Software Engineer)  
\end{quote}

To evaluate the simulator, the development team applied the LLM-as-a-Judge framework to assess how well the simulated instructor generated questions that reflected those an instructor might naturally ask. A team member developed tailored evaluation criteria to guide these automated assessments, which focused on dimensions such as pedagogical relevance, cognitive depth, contextualization, and breadth of instructor concerns. These automated evaluations suggested that the instructor dialogues ranged from good to excellent in quality, giving the developers confidence that the simulator could be well-suited to ask questions. 

In summary, each shift in data collection reflected a decision shaped by practical constraints. Recordings were abandoned because privacy concerns outweighed potential benefits, staged conversations proved too limited in size and format, and the simulator was adopted as a feasible way to involve experts more widely and build a usable dataset. The developers had to acknowledge potential issues and amend their approach to work around the issue of collecting expert feedback.


\subsubsection{Developers Turned to Augmentation Strategies Due to Limited Data Collected}

\revision{Upon reviewing the gold-standard dialogues collected through expert interactions with the simulator, the team determined that while the data were useful for evaluation, the sample was too limited in size and diversity to support fine-tuning.}
\delete{In reflecting on the set of gold standard dialogues collected from the experts interacting with the simulator, the team recognized that while these data were valuable for evaluation, it was too small a sample set to support fine-tuning. Through discussions, the development team concluded that the limited number of conversations could not provide the scale or diversity needed for robust training.} This recognition motivated a shift in strategy. Rather than continuing to seek additional expert dialogues through the simulator, they decided to adopt data augmentation methods to artificially expand the dataset and approximate the kinds of expert-like dialogues needed for training. One participant who was part of the development team summarized this reasoning:

\begin{quote}
\textit{``I had the instinct that I wouldn’t have enough data to train a model...[after group discussion]...I should think of new ways [to expand data], like augmentation.''} (UX Project Manager)  
\end{quote}

To implement this approach, the team generated synthetic multi-turn dialogues that modeled the expert data that had been collected from the simulator. The augmented conversations were crafted to mirror the structure and tone of authentic consultations while expanding the dataset to 1,000 synthesized dialogues. This step reflected the team’s focus on efficiency, as augmentation allowed them to quickly scale beyond what could be collected from experts in real time. In addition, the team did not conduct a direct evaluation of the augmented data itself due to time constraints and assessed the model’s augmented outputs only after fine-tuning (see Section \ref{Evaluation_methods}). 

In summary, augmentation enabled the developers to move forward despite limited data from experts to produce a larger and more consistent dataset for training. This shift, however, also marked a turning point in the design process: rather than grounding the model primarily in expert-provided conversations, the team chose to use synthetic data to approximate expertise at scale. 



\subsubsection{Developers Collaborated with Pedagogical Experts to Co-Develop Evaluation Criteria}

After generating data for fine-tuning the model, the team needed a way to evaluate whether the chatbot’s responses reflected sound pedagogical practice. While domain experts offered the highest standard of evaluation, the team recognized that relying exclusively on them would be prohibitively time-consuming and costly. Instead, they began exploring strategies that could balance the domain experts' oversight with scalable evaluation methods.

Within this process, the idea of using an LLM-as-a-Judge again emerged as a central strategy to capture the essence of expert judgment without requiring experts to rate every output. Their solution was to translate expert feedback into evaluation criteria that the LLM-as-a-Judge could apply at scale. This reasoning reflected a compromise: experts were still essential for setting standards, but once those standards were formalized, the automated judge could take over routine evaluation. In this way, the team preserved the authority of expert input while creating a workflow that was more efficient, sustainable, and repeatable. As one developer explained, the team’s limited pedagogical expertise meant they could only achieve a baseline score with the LLM-as-a-Judge alone, but expert guidance could significantly raise the benchmark:

\begin{quote}
    \textit{``We are not experts in this domain...we have a baseline [set of criteria] that will be rated [by the LLM-as-a-Judge] six out of 10, but we can get to probably nine out of 10 [with expert input into defining criteria]''} (UX Project Manager)  
\end{quote}

This criteria was used for two main purposes: first, to quickly evaluate the model and identify outputs that did not align with expert expectations, and second, to serve as metrics for comparing the fine-tuned model to a standard GPT model and testing performance differences.

Initially, to create criteria, the development team drew on existing standard practices for pedagogical consultation without experts to draft an initial set of criteria grouped into categories. These criteria were created from utilizing online resources on ``best practices'' for assessing both the quality of pedagogical advice and the overall coherence of the dialogue. However, the developers realized that the criteria that they developed were generic and lacked the domain specificity that they could not supply themselves. As one developer explained:

\begin{quote}
    \textit{``We didn't want to use standard evaluation criteria, such as [measures of] fluency or typical checks of how correct the language is... those are criteria that can be used for every LLM task. It is important that we trust human involvement more.''} (UX Project Manager)  
\end{quote}

With this in mind, the team concluded that the most effective way to elicit high-quality input from domain experts was to organize a series of participatory design workshops, rather than simply asking individuals to generate evaluation criteria independently. They reasoned that a collaborative process would make it easier for experts to refine their ideas, align on shared standards, and better connect their knowledge to the system's needs. As one developer explained, when thinking about the time and cost constraints: 

\begin{quote}
    \textit{``I feel like the only way that we can get experts [to actively participate] is to do a workshop so that they can focus and get the thing done.''} (UX Project Manager)
\end{quote}

The team intentionally recruited pedagogical experts with diverse backgrounds, such as consultants who regularly advised faculty, researchers who studied pedagogy, and professors with practical classroom experience to surface different perspectives on what counts as ``good'' teaching advice. The development team conducted three small-group sessions, each lasting 1.5 hours, with two \revision{hourly compensated} experts to facilitate in-depth discussions. \delete{Experts were compensated hourly for their contributions.} In planning these sessions, the developers also debated how best to capture feedback, such as through whiteboard brainstorming, paper annotations, or digital criteria editing. Ultimately, they distributed the draft criteria in a shared document, allowing experts to directly revise and suggest edits on their laptops. \delete{This decision reflected the team’s desire to make feedback collection easy and efficient while lowering barriers for participants.}  

The workshops followed a structured progression, with each step intentionally designed to strike a balance between efficiency and expert understanding. First, experts were briefed on the project’s goals, the purpose of the evaluation criteria, and the concept of using LLM-as-a-Judge to scale assessment. The team reasoned that without this orientation, experts might view the task as a traditional human assessment exercise, rather than understanding how their criteria would be operationalized by an automated judge. Second, experts reviewed the draft criteria created by the developers individually and in discussion, and suggested amendments or additions. Third, experts were shown chatbot outputs alongside the LLM-as-a-Judge’s automated ratings and asked whether the criteria were capturing meaningful qualities. This comparison was included to make the abstract concept of automated evaluation more concrete, giving experts a chance to see how the judge ``interpreted'' the initial criteria in practice. Finally, experts interacted directly with the chatbot to test whether their refinements better reflected pedagogical standards in real-world use. This final step created an opportunity for iterative reflection and ensured that the criteria were not only theoretically valid but also practical in guiding model evaluation.

\revision{During the initial rubric review, the experts' criteria evolved to be more descriptive, particularly by clarifying the guiding statements or questions used by the LLM-as-a-Judge. As experts compared chatbot outputs with LLM-as-a-Judge scores, they identified disagreements between their own assessments and the judge’s rankings. These mismatches prompted attempts to clarify wording, refine numerical scales, and add contextual guidance. However, because changes to the criteria could not be tested interactively to observe immediate updates in LLM-as-a-Judge scoring, refinements remained limited to minor edits rather than substantive restructuring. After experts interacted directly with the chatbot, no major structural changes to the criteria were introduced, though several usability issues were identified and documented by the development team.}

In summary, the developers recognized their own limitations in defining robust evaluation criteria and sought expert involvement to ensure that the chatbot was assessed against pedagogical standards of quality and accuracy. \delete{At the same time,} The team also invested considerable effort in designing strategies for introducing criteria to present them in ways that would allow experts to readily engage with, critique, and refine them.


\subsubsection{Developers Used a Hybrid Evaluation Approach to Address Time Constraints and Inconsistencies}
\label{Evaluation_methods}
After fine-tuning the model with new augmented training data and establishing evaluation criteria, the development team adopted a hybrid evaluation approach that combined domain expert input, developer judgment, and automated scoring from the LLM Judge. 

Initially, the team asked pedagogical experts to evaluate chatbot responses using the criteria developed in earlier sessions, with the goal of comparing expert ratings to those produced by the LLM-as-a-Judge. To do this, experts participated in informal interviews in which they reviewed chatbot responses and ranked the outputs using the criteria. However, this task quickly proved too tedious, as experts had to read each full conversation, decide which criteria applied, and provide both a rating and justification. 

The team then shifted responsibility to members of the development group, who were more readily available to conduct evaluations. Developers reasoned that, after working closely with experts in participatory design sessions, they had developed a reasonable sense of how experts would apply the criteria. This gave them confidence to take over the evaluations in order to save time and flag any obviously erroneous responses.

In addition, the team used the LLM-as-a-Judge to rank the provided criteria against both their chatbot and a different GPT model in order to compare performance. \revision{To do this, they tested different LLM-as-a-Judge prompts until the rankings aligned more closely with expert and developer manual assessments.} Yet development team members noted that automated scores were not always consistent:

\begin{quote}
\textit{``If we use the LLM-as-a-Judge, sometimes the scores are not consistent. Without [the LLM] giving a justification, it's kind of a black box. For example, if I set the temperature to zero, the scores may be consistent, but if I add something else to the prompt, the scores can change.''} (Technical Lead)  
\end{quote}

Despite these limitations, developers would discuss which criteria were more impactful, meeting regularly to review results and decide whether metrics truly differentiated model performance. As one developer described, the process was less about finding perfect scores than about identifying which criteria meaningfully discriminated between the models that they were comparing:

\begin{quote}
\textit{``We look at how the scores turn out. If most of the scores are perfect, maybe that criterion is not important. Then we try another criterion to see if one model scores differently from another, which can help show meaningful distinctions between models.''} (Technical Lead)  
\end{quote}

These findings suggest that while the development team recognized expert input as ideal, they ultimately relied on developer involvement and automated LLM-as-a-Judge methods to conduct evaluations. Importantly, the criteria not only shaped how evaluations were conducted but also informed model refinement. Developers used patterns in the scores, such as criteria that consistently exposed weaknesses, to guide adjustments in fine-tuning and prompt engineering and effectively looped evaluation results back into the development process.

\subsection{Challenges and Trade-offs  \revision{(RQ2)}}

Five overarching themes were identified concerning the challenges and trade-offs the team faced as they navigated the design and evaluation of the LLM chatbot.


\subsubsection{Motivations and Trust Featuring Expert Participation}

Interviews with pedagogical experts revealed that their willingness to participate was shaped by several motivating factors, including professional curiosity about AI, institutional trust, available time, and compensation. For some, the experience was described as intellectually and professionally rewarding. Others emphasized a desire to be involved so that they could better understand the tool and eventually promote it to faculty instructors:

\begin{quote}
    \textit{``It gives a lot of credibility if I'm going to be working with a faculty member, and say, hey, there's this tool it's pretty helpful, I know the designers, and here's what goes into it, as opposed to some company somewhere made something.''} (Domain Expert 3)
\end{quote}

Another expert described initial hesitation toward participating, stemming from uncertainty about AI and its implications for writing instruction. This reluctance evolved into motivation as participation became a chance to learn, reflect, and contribute meaningfully to the project:

\begin{quote}
\textit{``I was a little reluctant at first because I don’t use AI...but I always walked away from sessions feeling like I’d learned something new...It was definitely more work than I expected, but the time felt well spent.''} (Primary Stakeholder)
\end{quote}

Another domain expert framed skepticism as a productive motivation for engagement, emphasizing that diverse viewpoints strengthened the design and evaluation process:

\begin{quote}
    \textit{``I appreciate being included because I am an AI skeptic. I support the use of AI, and I think we should use and embrace it, but I feel like I'm often the most skeptical person in the room in a lot of meetings...it's good to have people with a range of opinions about AI.''} (Domain Expert 2)
\end{quote}

Trust in the institution emerged as a consistent theme, as experts felt comfortable contributing because the project was led by colleagues within their university rather than driven by profit motives. This sense of institutional alignment made participation feel safer and more meaningful. Experts stated that they would have felt less comfortable contributing their expertise to commercial or for-profit efforts:

\begin{quote}
    \textit{``In an organization that has thoughtful leaders, that has strong consensus around institutional values and goals and adequate resources to support all of that...I'm not worried .''} (Domain Expert 3)
\end{quote}

In summary, the pedagogical experts emphasized that their motivation to participate was not only practical, as driven by their curiosity, trust, or compensation, but also ethical. Experts carried with them a sense of responsibility for how their knowledge would be used and expressed a desire for their contributions to benefit the university community rather than external commercial interests. \delete{This framing shows that expert involvement was shaped as much by professional values and ethical standards as by logistical factors.}



\subsubsection{Challenges in Structuring Participatory Design Processes}

Observations of the participatory design sessions revealed several challenges that shaped how domain experts engaged with creating evaluation criteria for the LLM-as-a-Judge. While the workshops provided valuable input, experts frequently struggled to distinguish between evaluating the chatbot itself and refining the criteria. These challenges stemmed from three main factors: limited prior experience with design sessions, difficulties in understanding the LLM-as-a-Judge concept, and insufficient time for reflection.

\paragraph{Limited Experience with Design Sessions. } For many experts, this was their first time participating in a structured design session to create evaluation criteria. The process involved reviewing model outputs and comparing them to criteria, which included tasks that felt unfamiliar and cognitively demanding. As one expert reflected:

\begin{quote}
\textit{``I've never done anything like that before...so it was challenging for me to remember the steps that the LLM was evaluating the chatbot, and we were evaluating the criteria. It took me a minute to figure out what we were actually doing or what we were actually evaluating.'' (Domain Expert 5)}
\end{quote}

Several experts described the process as ``intense'' and suggested that receiving the criteria beforehand might have made the task more manageable:

\begin{quote}
\textit{``I think the process was a bit intense...For me, it takes time to connect the evaluation criteria to my work experience...so I need more time to process... maybe it will be better if I can access these evaluation criteria beforehand to make more connections.'' (Domain Expert 4)}
\end{quote}

\paragraph{Difficulties Understanding the LLM-as-a-Judge Concept. } Although the development team introduced the idea of the LLM-as-a-Judge, some experts found it difficult to keep this framework in mind during the sessions. They often defaulted to thinking about how they would want the chatbot to behave, rather than focusing on whether the evaluation criteria captured meaningful standards. As one participant explained:

 \begin{quote}
    \textit{``It was tricky, because I was thinking to myself about what I would want out of [the chatbot], but then I also had to make that extra step of thinking that I'm looking at the evaluation criteria...just like trying to remind myself to stick with the LLM Judge piece of it. I think we got there eventually.'' (Domain Expert 5)}
\end{quote}


\paragraph{Insufficient Time for Reflection. } Experts valued the collaborative environment of the sessions but noted that more time to reflect on the criteria would have improved the process. They suggested follow-up opportunities or real-time iteration to better connect the criteria with practice. One expert described the benefit of live feedback loops:

\begin{quote}
    \textit{``Instead of doing [the criteria edits] on paper...do it in real time....make changes live. [Such that] we suggest some language or a few descriptors to the criteria, then we look at the output...and see what does it rate...to see how that's changing...[whether] it's not interpreting the information the way an expert would.''} (Domain Expert 3)
\end{quote}

These findings show that while participatory design workshops offered an opportunity to align evaluation criteria with expert standards, they also opened discussion around the use of expertise, the understanding of AI, and time. Experts expressed the need for more guidance and opportunities for reflection to fully engage and best contribute to the process.

\subsubsection{Concerns About How Expert Feedback Would Be Used}

While many experts described their participation as rewarding, they also voiced concerns about how their contributions would ultimately be used in the development of the LLM chatbot. \delete{Several participants noted that they were effectively sharing their ``tips and tricks'' of consultation practice, which raised questions about ownership of this knowledge once it was embedded into the system.} This was mostly about sharing data as well as the ability of the chatbot to mimic the job of an expert.

\paragraph{Concerns About Giving Away Knowledge. } Experts had some reservations about contributing their expertise to the project and how it might be repurposed. While they recognized that providing data to the simulator could help train novice consultants by modeling realistic interactions, imputing their tacit ``tips and tricks'' into a dataset raised concerns about ownership and loss of control once the knowledge was embedded into the system. As one expert reflected:

\begin{quote}
    \textit{``I was working with the simulator...even though you're trying to take my job physically...I'm giving it all my tips and tricks on how I go about my consultation.''} (Domain Expert 3)
\end{quote}

\paragraph{Concerns About Replicating Expert Work. } Most domain experts were confident that the LLM could not replicate the full scope of their work. They emphasized that pedagogical consultations often involve faculty sharing personal or vulnerable information, which requires confidentiality, trust, and professional judgment to handle responsibly. These dimensions go beyond providing advice and reflect the ethical and relational aspects of expert practice. As one expert explained:

\begin{quote}
    \textit{``We try to protect the confidentiality of our clients...there are times when faculty come to us in a place of vulnerability and we would like to be able to help that person without it putting their career in jeopardy.''} (Domain Expert 1)
\end{quote}

These reflections suggest an important trade-off in expert involvement. Expert knowledge can either be essential to creating meaningful evaluation criteria and training data, or it could surface anxieties about ownership, control, and the irreplaceable aspects of human reasoning and care in pedagogical practice.



\subsubsection{Fatigue and Engagement in Data Contribution}

The simulator enabled the team to collect dialogues at scale; however, experts reported that the process was repetitive and, over time, exhausting. \delete{Repeated prompts often limit variety, making it difficult to sustain focus.} Experts developed their own strategies for staying engaged, such as selectively choosing more interesting or challenging scenarios. As one expert explained:

\begin{quote}
\textit{``I would try to do 15 to 20 minutes [of the conversations]...I could only get through two or three at a time before some of the prompts or problems that it presents are kind of repeated throughout...so I try to find one that's more interesting or different to keep myself engaged right in the process''.} (Domain Expert 3) 
\end{quote} 

Beyond repetition, experts also highlighted that the typing-based format of the simulator created additional strain. They suggested that alternative modes of contribution to the simulator, such as using voice-to-text, could help sustain longer engagement. As one expert noted:

\begin{quote}
\textit{``If there's an opportunity for a voice interaction in conversation...that would certainly sustain longer attention and engagement with it, because it's similar to a conversation with the person.''} (Domain Expert 3) 
\end{quote} 

These findings reveal that while the simulator made expert involvement more scalable, the format could introduce trade-offs in terms of fatigue and depth of contributions. Designing future workflows may require diversifying interaction formats to reduce repetition and better mirror the natural dynamics of consultation practices.

\section{Discussion}
This study provides an in-depth examination of how expert-informed evaluation unfolds within an academic LLM development team. Our findings reveal both the potential and complexity of involving domain experts in the design and evaluation of AI systems. While the development team valued expert input and experts were motivated to participate, resource constraints, mismatched participation formats, and limited AI literacy support introduced challenges. These frictions point to the need for intentional design of workflows, tools, and ethical practices that can better support meaningful and sustainable expert engagement. In this section, we synthesize our findings into three key lessons for HCI researchers and practitioners seeking to integrate domain expertise into LLM development \revision{(RQ3)}.







\subsection{Designing Tools That Capture Rich Expert Knowledge}

Recent work in ML has increasingly emphasized data-centric approaches to improving model quality~\cite{jakubik2024data}. Our findings extend this perspective by showing how the design of data collection tools fundamentally shapes the kind of expert knowledge that can be captured~\cite{pan2022data}. In our study, the simulator served as a flexible data collection tool and gave experts the autonomy to contribute at their own pace. However, the repetitive questions and decontextualized prompts constrained their ability to express the deeper reasoning that reflects real consultations. This raises questions about whether the tool adequately supported the diversity of consulting scenarios and instructional queries relevant to expert practice.

Prior research highlights the persistent difficulty of encoding professional expertise for computational systems~\cite{pakarinen2025relational}. In our study, experts noted that short written responses in the simulator did not fully express deeper pedagogical reasoning, which risked reducing their expertise to surface-level data. This aligns with critiques in the ML community that model development often prioritizes algorithmic performance over the quality and representativeness of training data~\cite{whang2023data, inel2023collect}. When AI systems aim to supplement or replicate expertise, the risks of shallow data become especially concerning, since insufficient training data may harm users~\cite{szymanski2025limitations}.

\textit{Lesson 1: Experts should be provided with tools that are both flexible and data-centric for collecting insights.} 
Future workflows should consider how to elicit richer forms of expert knowledge. Approaches such as interactive demonstrations, think-aloud protocols, or focus sessions that support reflection may help capture not only explicit ``answers'' but also the strategies and reasoning processes underlying expert practice. In doing so, data-centric design tools can better balance the needs of developers for structured datasets with the needs of experts for autonomy in representing their work.




\subsection{Evolving Roles of Experts in the Era of AI}
Domain experts often struggled with a limited understanding of AI concepts, which made it challenging to fully grasp the task of co-designing evaluation criteria. Although they possessed the necessary knowledge and experience to contribute to the system evaluation, the experts struggled to fully grasp how their knowledge was being applied in the LLM-as-a-Judge, even after being introduced to the concepts and the co-design task. This raised two key risks: first, that the evaluation criteria might result in reduced quality or misalignment with the system’s needs; and second, that experts could unknowingly contribute knowledge without fully approving of its usage and application within the evaluation.

This finding is aligned with other literature that supports incorporating AI literacy support into participatory design~\cite{dangol2024mediating}. Previous studies indicate that stakeholders are commonly engaged in critiquing system interfaces even though they are not properly prepared to assess or understand the underlying model itself~\cite{delgado2021stakeholder}. Research has also shown that all participants need certain design skills and conceptual understandings to be most successful in these settings~\cite{qi2025participatory, zytko2022participatory}. Our study reinforces this gap and provides further evidence that structured AI literacy support is essential for domain experts to make meaningful contributions to the development and evaluation of AI systems.

Beyond completing the task, experts should be able to critically question the goals of the system, the risks of automation, such as LLM Judge approaches, and the boundaries of their own contributions. This requires creating reflective spaces in the design process, rather than just providing task instructions. Long et al. have discussed specific core competencies required for AI literacy~\cite{long2020ai}, including high-level overviews of how algorithms operate, explainable AI features that clarify outcomes, and interactive demonstrations that enable experts to test or simulate system behavior, all of which contribute to AI literacy~\cite{long2020ai}. In our study, for instance, domain experts only began to fully grasp the role of the LLM-as-a-Judge once they were shown sample scores accompanied by the judge’s explanations. However, what was missing from the team's participatory design process was a mechanism for experts themselves to test and refine the criteria in real-time. One expert explicitly discussed the need for interactive tools that support the iterative development and refinement of criteria within the design session. Recent HCI work has explored interactive evaluation support tools (e.g.,~\cite{gebreegziabher2025metricmate, kim2024evallm, desmond2025evalassist}) which can be used to support all stakeholders in these evaluation contexts.

Our findings also point to broader shifts in the nature of expert labor. As domain experts in this study were increasingly asked to train and evaluate AI systems, their consulting roles expanded from advising instructors to shaping algorithmic decision-makers. This shift signals a potential redefinition of professional identity, from a consultant to a data provider or even a co-designer. 

\textit{Lesson 2: Expert Involvement Requires Both AI Literacy Support and Recognition of Shifting Roles.} HCI researchers and practitioners should take seriously the implications of this evolution. Sustaining expert engagement requires not only technical support for AI literacy but also structural recognition of experts’ changing roles and contributions in the era of AI. The involvement of experts should be considered a two-way exchange, as they need sufficient training and support to understand how their knowledge is being used, and at the same time, their labor must be valued as more than just data collection. As LLM Developers recruit domain experts for their tasks, they should be explicit about the roles needed and provide proper training.

\subsection{Ethical Implications of Expert Contributions}
Involving domain experts in LLM development introduces not only technical advantages but also broader ethical considerations related to how expert knowledge is incorporated and sustained over time. Pedagogical experts were asked to share their expertise by contributing ``tips and tricks'' for consultations through the simulator as well as refining evaluation criteria to reflect their best practices. Most of these experts had spent years developing their skills through training and professional experience. As the project progressed, expert-generated materials were incorporated into multiple stages of the development pipeline, including data augmentation and model fine-tuning, in ways that supported evolving technical and resource needs. While these practices were consistent with approved data use and project goals, the rapid integration of expert knowledge into automated workflows meant that contributions became increasingly abstracted from the individuals who produced them. This observation points to the importance of planning and governance mechanisms for data reuse that support transparency and help preserve expert agency as development priorities and system capabilities evolve.

This raises an important question: \textit{who owns a domain expert’s data once it is embedded into an AI system?} This concern is not unique to pedagogy~\cite{malhotra2024owns}. Broader debates are emerging across creative and professional domains, with artists discovering their work in image-generation datasets~\cite{zhang2025rise}, authors noting that their books have been scraped to train large models~\cite{Asmelash_2023}, and musicians raising concerns about voice synthesis and imitation~\cite{Team_2025}. In each of these cases, once expertise or creative labor is transformed into data, it becomes difficult to trace, attribute, or govern. Our findings suggest that the same problem extends to professional expertise. Pedagogical knowledge, developed through years of practice, risks being reduced to invisible ``training material'' for LLMs. This echoes concerns raised in Ghost Work about the invisibility of labor that sustains AI systems~\cite{gray2019ghost}.

This invisibility raises critical questions about ownership and recognition. When inventors patent a process, they receive recognition, legal protection, and sometimes royalties. By contrast, domain experts who provide strategies, evaluation heuristics, or consultation practices may receive short-term compensation (e.g., hourly pay) but often lack recognition or control over how their contributions are utilized. If their knowledge becomes central to a system that is later scaled, commercialized, or deployed, they have little agency over its usage. \revision{Moreover, when real-world data is constrained by privacy, access, or scale, teams often rely on augmented or synthetic data, as observed in our study, further complicating questions of ownership. Prior work shows that such data can introduce bias or misrepresentation if not carefully governed~\cite{de2024recommendations}.}

\textit{Lesson 3: Experts Should Be Explicitly Informed About How Their Knowledge Will Be Used.} While research teams cannot reshape global data policy, they can implement more transparent data collection practices. Recruitment and consent forms should state how expert contributions will be utilized, whether for training, evaluation, or deployment. Teams should also consider adopting licensing-style agreements that define boundaries of use (e.g., restricted to research, restricted to this organization's use) and clarify ownership. In addition, compensation structures could be expanded to recognize the long-term value of knowledge contributions, particularly when these shape evaluation frameworks or fine-tuned models. This might include authorship, acknowledgment, or shared rights to future system outputs. \revision{When synthetic data or simulators are used in place of real expert-authored data, experts should be involved in reviewing and refining generated examples over time to mitigate risks and retain influence as systems evolve.} Above all, systems and workflows should be designed to sustain expert agency \revision{through recurring expert review checkpoints} and ensure that experts remain collaborators and co-designers rather than invisible data providers.

\section{Limitations and Future Work}
Our work has limitations that also open directions for future research. First, this study focused on a single LLM development team observed over a 12-week period. Therefore, the findings reflect practices shaped by a specific institutional and organizational context and may not generalize to other domains or industry settings. 

Second, although multiple pedagogical specialists were included, all participants came from a single institution and shared similar professional backgrounds. Future work should broaden its scope to include experts from diverse disciplines, institutions, or cultural contexts, thereby enhancing understanding of how different forms of expertise influence LLM design and evaluation practices.

Third, the 12-week observation window captured only one phase of model refinement and evaluation and did not extend into deployment or long-term system use. As such, the study cannot address how expert–developer relationships or evaluation practices evolve over time. Future longitudinal research could investigate how collaboration practices, evaluation workflows, and perceptions of trust and ownership evolve as systems mature and are implemented in real-world contexts.


\section{Conclusion}
Through a twelve-week ethnographic study and interviews with developers and pedagogical experts, \revision{we investigated how an LLM development team approached the design and evaluation of a pedagogical chatbot, while navigating the challenges of integrating domain expertise.} \delete{Using ethnographic methods, we revealed} \revision{Our findings show} how the team navigated organizational constraints and limited expert availability through strategies, including data-collection workarounds, co-developing evaluation criteria, and hybrid expert–LLM assessment methods. \revision{These findings demonstrate the practical and ethical complexities of expert involvement, including issues of trust, motivation, and ownership.}
\delete{Our findings demonstrate the practical and ethical complexities of integrating domain expertise into LLM development provide insights into how design processes can better support expert participation and address common concerns around trust, motivation, and ownership.} For the HCI and IUI communities, this work offers design implications for workflows and tools that better support expert participation in LLM development. We argue that future intelligent systems should incorporate mechanisms for transparent consent, enhanced AI literacy, and recognition of expert contributions that allow domain specialists to participate as co-designers rather than solely as data providers. Such approaches can support more responsible, participatory, and sustainable LLM development practices.


\section{GenAI Usage Disclosure}
LLMs played a central role in the development activities observed in this study. These tools were employed by study participants as part of their normal development workflow. Generative AI tools (e.g., OpenAI’s GPT models) were also used to improve grammar and clarity in writing. All text was reviewed and verified by the human authors.

\begin{acks}
This work was supported in part by the Notre Dame–IBM Technology Ethics Lab. Any opinions, findings, and conclusions or recommendations expressed in this material are those of the authors and do not necessarily reflect the views of the sponsors.
\end{acks}

\bibliographystyle{ACM-Reference-Format}
\bibliography{base.bib}

\end{document}